# Experimental Report on Setting up a Cloud Computing Environment at the University of Bradford

Bashir Mohammad (b.mohammed1@bradford.ac.uk) School of Electrical engineering and computer science, Bradford University BD7 4DP

Mariam Kiran (m.kiran@bradford.ac.uk) School of Electrical engineering and computer science, Bradford University BD7 4DP

**Abstract**

Cloud computing is increasingly attracting large attention in computing both in academic research and in industrial initiatives. Emerging as a popular paradigm and an attractive model of providing computing, information technology (IT) infrastructure, network and storage to large and small enterprises both in private and public sectors. This project was initiated and aimed at designing and Setting up a basic Cloud lab Testbed running on Open stack under Virtual box for experiments and Hosting Cloud Platforms in the networking laboratory at the University of Bradford.

This report presents the methodology of setting up a cloud lab testbed for experiment running on open stack. Current resources, in the Networking lab at the university were used and turned into virtual platforms for cloud computing testing. This report serves as a practical guideline, concentrating on the practical infrastructure related questions and issues, on setting up a cloud lab for testing and proof of concept. Finally the report proposes an experimental validation showing feasibility of migrating to cloud.

The primary focus of this report is to provide a brief background on different theoretical concepts of cloud computing, particularly virtualisation, and then it elaborates on the practical aspects concerning the setup and implementation of a Cloud lab test bed using open source solutions. This reports serves as a reference for institutions looking at the possibilities of implementing cloud solutions, in order to benefit from getting the basics and a view on the different aspects of cloud migration concepts.

# Contents

# Introduction

Virtualization can be defined as a process of creating a virtual version over the actual version of a present infrastructure.[4] This may include, and is not limited to, a virtual computer hardware problem, operation system (OS), storage device, or computer network resources.

Virtualisation is simply the basic act of decoupling an infrastructural service from the physical assets on which that service operates. These services which include Compute, or Network are described in a data structure, and exists entirely in a software abstraction layer reproducing the service on any physical resource running the virtualization software as illustrated in the figure 1 below comprising of Virtual machines (VMs), virtual servers and physical servers. Operation System virtualization, server virtualization, application virtualization, storage virtualization, and network virtualization are examples of types of virtualization[5]–[7].

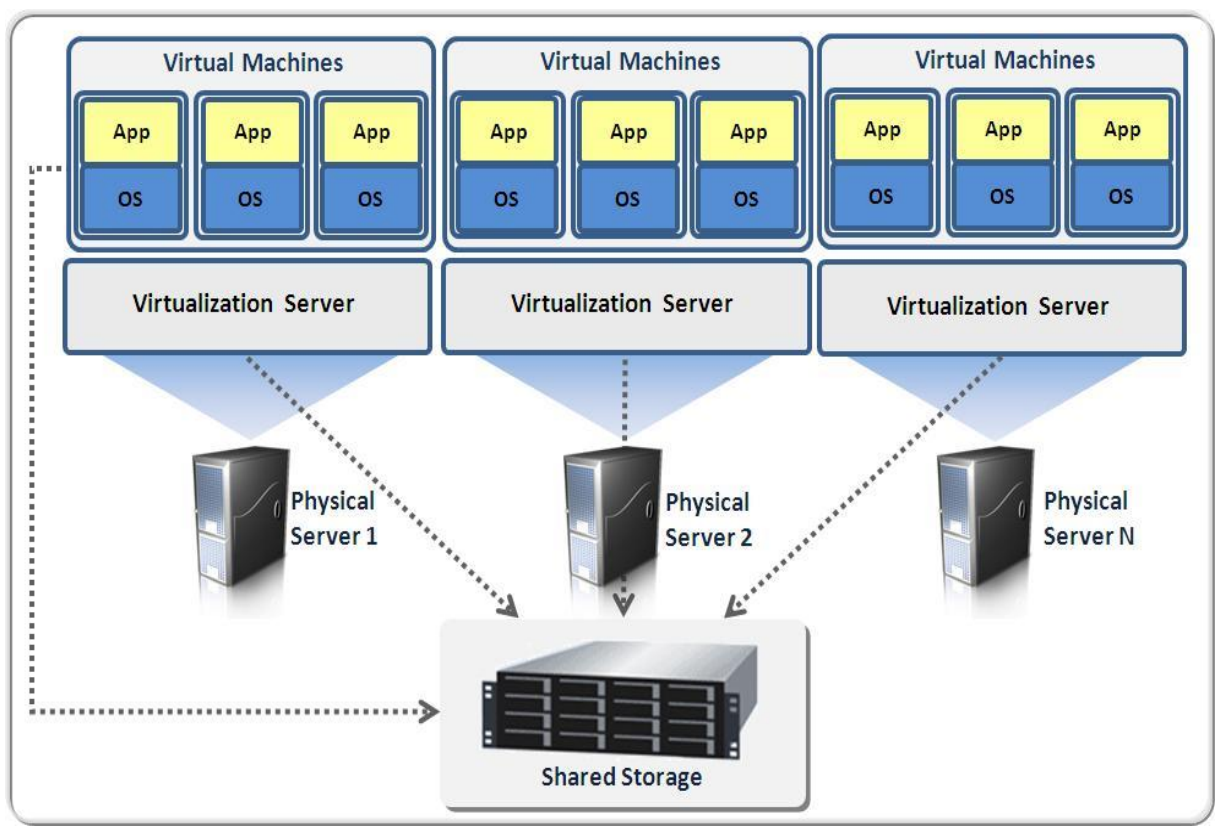

Figure 1: illustration of the concept of Virtualization [7]

The development virtualization technology was first developed in the 1960s on IBM mainframe[8]. The technology enables multiple virtual machines with guest operating systems to run simultaneously and independently, on a physical machine.

In the late 1990s, a breakthrough on x86 Platforms occurred, with some numerous virtualization projects been originated and developed. Some of the most popular virtualization products include VMware ESX Server, VMware Workstation, Parallels Virtuoso, Microsoft Virtual PC, and Microsoft HyperV, Sun xVM VirtualBox, QEMU, KVM and Xen[5].

Processors supporting hardware assisted virtualization are very important and this has been developed by Intel and AMD. The development of the services of cloud computing is a general model for delivering information technology (IT) services, on demand, over a private or public network have been facilitated by improvements in virtualization and distributed computing. It is a well-known fact that the reduction of hardware and maintenance costs can be improved with the aid of Virtualization and cloud computing technologies[9]. This can also improve the availability of resources, and expedite the deployment of new services. Adoption of the technology has been widely accepted in the industry.

# Background

### VIRTUAL BOX

Oracle VM VirtualBox (formerly Sun VirtualBox, Sun xVM VirtualBox and innotek VirtualBox), a virtualization software package for x86 and AMD64/Intel64-based computers from Oracle Corporation, forms part of Oracle's family of virtualization products[10]. Innotek GmbH first developed the product and Sun Microsystems purchased it in 2008 then Oracle has continued an active development since 2010[5], [11].

The Virtual Box package installs on an existing host operating system as an application; this host application allows additional guest operating systems, each known as a *Guest OS*, to load and run, each with its own virtual environment. Supported host operating systems include Linux, Mac OS X, Windows XP, Windows Vista, Windows 7, Windows 8, Solaris, and Open Solaris; there are also ports to FreeBSD and Genode[11].Supported guest operating systems include versions and derivations of Windows, Linux, BSD, OS/2, Solaris, Haiku and others. Guest dditions should be installed in order to achieve the best possible experience [12].The Guest Additions are designed for installation inside a virtual machine after the installation of the guest operating system. They consist of device drivers and system applications that optimize the guest operating system for better performance and usability[13].

### OPEN STACK

Open Stack, also called the open source cloud operating system, is a cloud operating system that controls large pools of compute, storage, and networking resources throughout a datacentre, all managed through a dashboard that gives administrators control while empowering their users to provision resources through a web interface[14], [15].

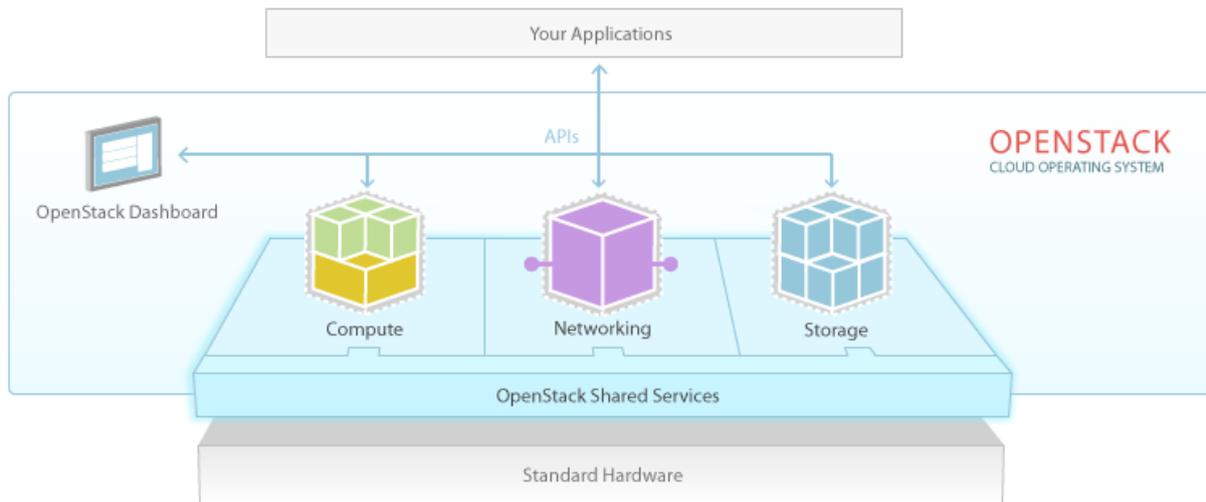

Figure 2:The Open Source Cloud Operating System overview [15]

Users primarily deploy the project as an infrastructure as a service (IaaS) solution. The technology consists of a series of interrelated projects that control pools of processing, storage, and networking resources throughout a data centre. Users can manage their cloud environment through a web-based dashboard, command-line tools, or a Restful API.

OpenStack Compute Provision's and manage large networks of virtual machines. The OpenStack cloud operating system enables enterprises and service providers to offer on-demand computing resources, by provisioning and managing large networks of virtual machines. Compute resources are accessible via APIs for developers building cloud applications and via web interfaces for administrators and users. The compute architecture is designed to scale horizontally on standard hardware, enabling the cloud economics companies have come to expect.

OpenStack is designed to provide flexibility as you design your cloud, with no proprietary hardware or software requirements and the ability to integrate with legacy systems and third party technologies. It is designed to manage and automate pools of compute resources and can work with widely available virtualization technologies, as well as bare metal and high-performance computing (HPC) configurations. Administrators often deploy OpenStack Compute using one of multiple supported hypervisors in a virtualized environment. KVM and XenServer are popular choices for hypervisor technology and recommended for most use cases. Linux container technology such as LXC is also supported for scenarios where users wish to minimize virtualization overhead and achieve greater efficiency and performance. In addition to different hypervisors, OpenStack supports ARM and alternative hardware architectures[14]–[16].

An example of some of the most popular cases of Openstack include service providers offering an IaaS compute platform or services higher up the stack, IT departments acting as cloud service providers for business units and project teams, Processing big data with tools like Hadoop, Scaling compute up and down to meet demand for web resources and applications, High-performance computing (HPC) environments processing diverse and intensive workloads[15].

## The Problem with Brief Literature Review

Recent studies have looked into the use cloud lab test beds for experiments. They have generally focused on demonstrating the efficiency of cloud migration and studying how the current resources can be turned into virtual platforms for cloud computing testing.

Nasim & Kassler 2014 [14] provided a comparative view on the performance of OpenStack while deploying it over a virtual environment versus using dedicated hardware. Two separate testbeds of OpenStack were deployed: one on virtual environment and the other on dedicated hardware. Each deployment was made with two computers where one computer acts as "Controller Node" and another one is used to provide computing service and treated as "Compute Node". Three basic tests were carried out on both environments to check CPU performance, data transfer rate, and bandwidth [14]. The overall aim of the experiments was to quantify the impact on performance of OpenStack when the underlying infrastructure is changed from virtualized to physical environment. Their results indicated that OpenStack over dedicated hardware performs much better than Open Stack over virtualized environment.

Adami et al. 2013 [17], investigated such issues in an experimental testbed, focusing on the measurement of the traffic overhead due to virtual machine migration in different operating conditions. Their measurement campaigns highlights that performance is strongly affected by several factors, such as VMs placement, VMs active memory. The result of their study suggested that the redistribution of VMs following the failure of a host or sudden increase of the required resources must be made taken into account the overhead the VM migration produces.

Alves & Prata 2014 [18] proposed a platform named CSC (Computer Science Cloud) Laboratory for the provisioning and management of laboratories with the computing resources necessary to teach a class using the cloud. The system was designed to be used by non-experts in cloud technologies and implemented to run in OpenStack but it can be easily adapted to other platforms with similar services. They presented a CSCLab, a proposal for a platform to manage computer science laboratories by using cloud resources trying to take advantage from the elastic cloud model. They also concluded that using the cloud, computing resources could be allocated or released according to a greater or lesser demand. This platform is intended to be used by non-experts in Cloud technology.

## The Experiment

### Testbed Set-Up
To set-up our experimental scenario, we started by setting up a testbed for hosting a Cloud platform. We opted for the OpenStack solution to enable us controls large pools of compute, storage, and networking resources throughout a datacentre on out testbed network [ 2 ]. This can all be all managed through a dashboard that gives administrators control while empowering users to provision resources through a web interface[14], [15].

The OpenStack solution is recent and under active development. It has great potential due to its architecture and large community and the support of its partners. All code is licensed under Apache 2 license. Openstack is supported by many companies in the world and is based on the code used by NASA and Rackspace Cloud. It is written in python and currently implements two control APIs, the EC2 API and Rackspace. It uses different drivers to interface with a maximum number of hypervisors such as Xen, KVM, HyperV, and Qemu [19], [20].

The testbed was composed of three hosts (See Figure 3)

| HOST | CPU | RAM | SYSTEM TYPE |
|---|---|---|---|
| 1 | Intel(R) Pentium(R) 4 CPU 3.20GHz 3.19 GHz | 2.00 GB | 64-Bit OS |
| 2 | Intel(R) Pentium(R) 4 CPU 3.20GHz 3.19 GHz | 2.00 GB | 64-Bit OS |
| 3 | Intel(R) Pentium(R) 4 CPU 3.20GHz 3.19 GHz | 2.00 GB | 64-Bit OS |

Table 1: Hardware Configuration of the DC Hosts

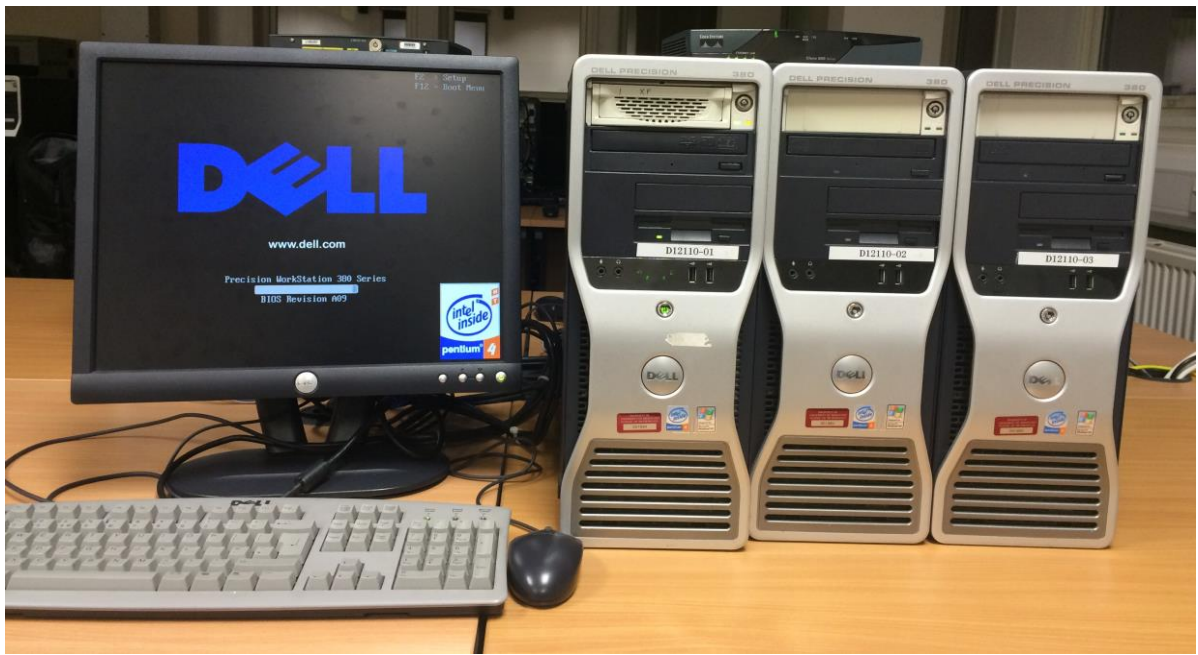

Figure 3: Experimental Setup at the Networks and Systems Lab University of Bradford

This project is an experimental study on how institutions will have the opportunity of building a hosting architecture and massive scalability, as it is completely open source, while it overcomes the constraints of the use of proprietary technologies.

### Installing Openstack
On the first three hosts described in table 1, we installed a Windows 7 Operating Systems with a 64-Bit operating system (OS) type. Inorder to test OpenStack, we installed Devstack, which is primarily used to build a complete Openstack development environment. It was downloaded from http://www.devstack.org

### The Operating Systems

This installation was done on a Linux (Ubuntu) OS. The recommended operating systems include Ubuntu, Fedora CentosOS/RHPL. As a requirement, the installation of DevStack required Virtual box which was downloaded from https://www.virtualbox.org. This is the virtualization software package for x86 and AMD64/Intel64-based computers from Oracle Corporation[1].

### Devstack Installation Process

The Devstack installation process carried out is summarized as follows:

1- Linux (Ubuntu) installation on Virtual Box (figure 4-6).
2- Basic upgrade on Linux using the following commands (figure 7)
    $ sudo apt-get update
    $ sudo apt-get –u upgrade
3- Install git (figure 7)
    $ sudo apt-get install git
4- Install the latest version of DevStack from GitHub
    $ git clone –b stable/icehouse http://github.com/openstack-dev/devstack.git
5- $ cd devstack
6- $ Creating localrc file
7- $. /stack.sh

Note: DevStack installation process leads to OpenStack dashboard shown in (figures 8-17)

During this process the following was assigned

ADMIN_PASSWORD=openstack
MYSQL_PASSWORD= openstack
RABBIT_PASSWORD= openstack
SERVICE_PASSWORD= openstack
SERVICE_TOKEN= tokentoken

The script allowed downloading dozens of dependencies, which are conveniently packaged by UBUNTU and the upstream Debian distribution. Setup.py for a few python packages is been run and some repositories and cloned.

8- NAT port forwarding
9- http://localhost to connect to Horizon

[Sample localrc file download] wget 0- localrc http://goo.gl/OeOGqL

[NAT port forwarding using iptables] sudo iptables –t nat –A POSTROUTING –o eth0 –j MASQUERADE

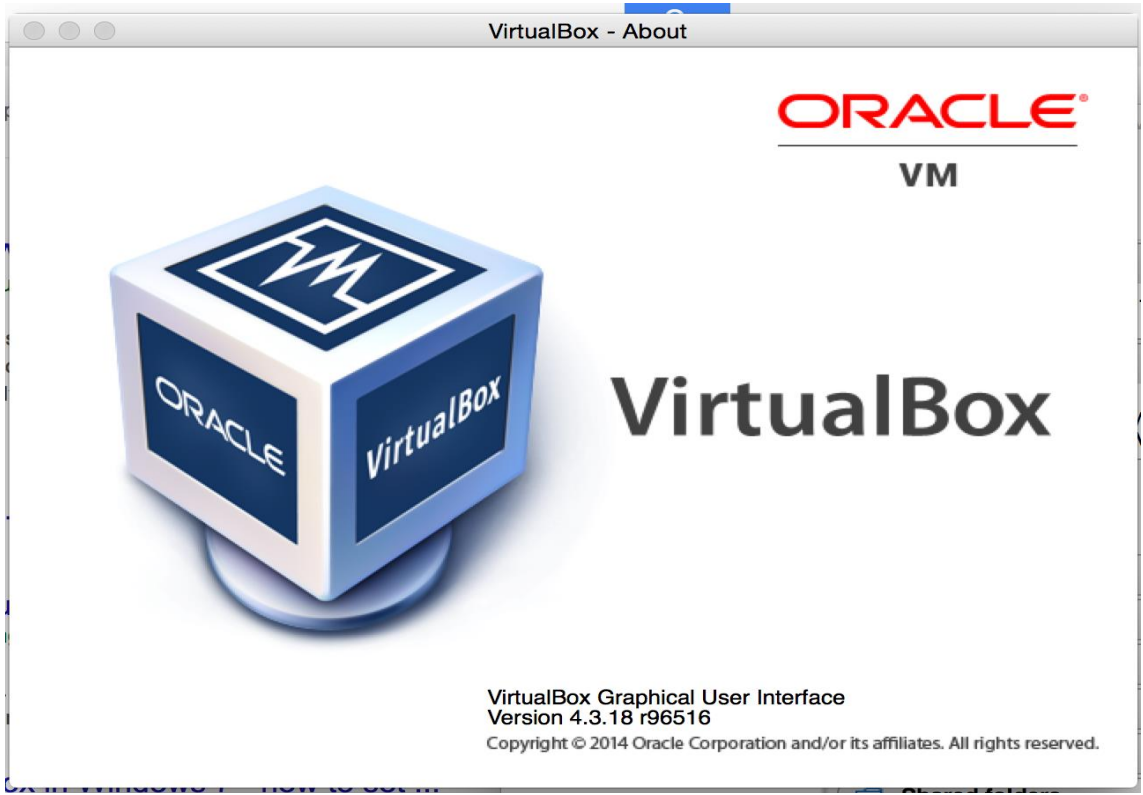

Figure 4: Virtual Box Installation start-up page

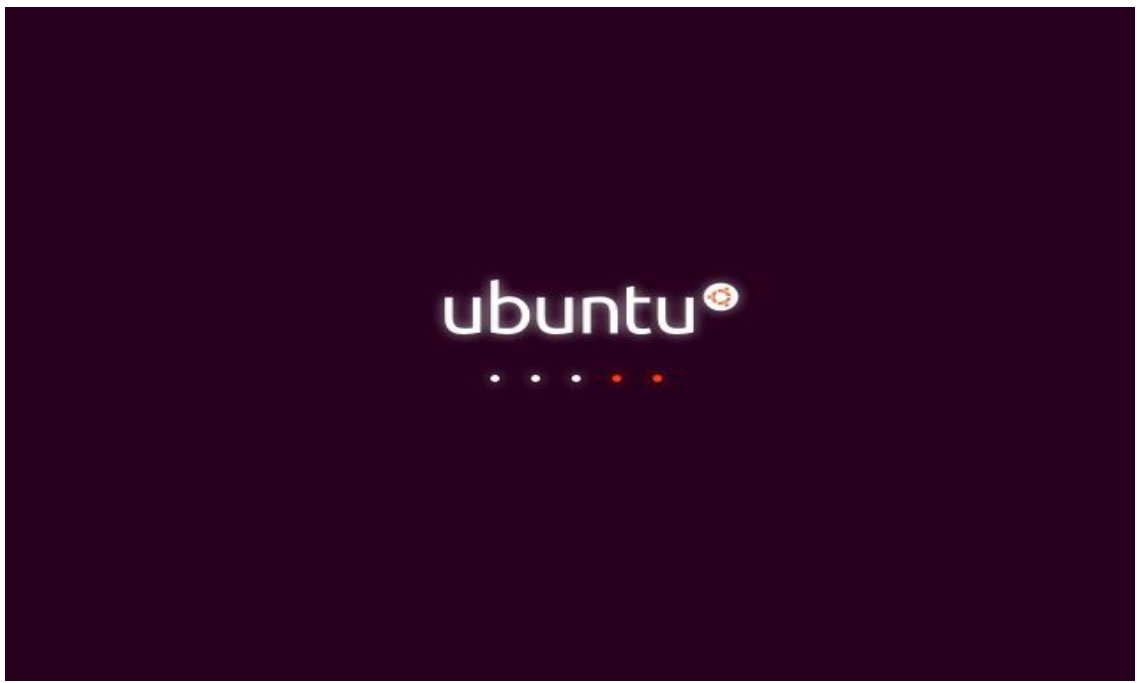

Figure 5: Ubuntu Installation start-up page

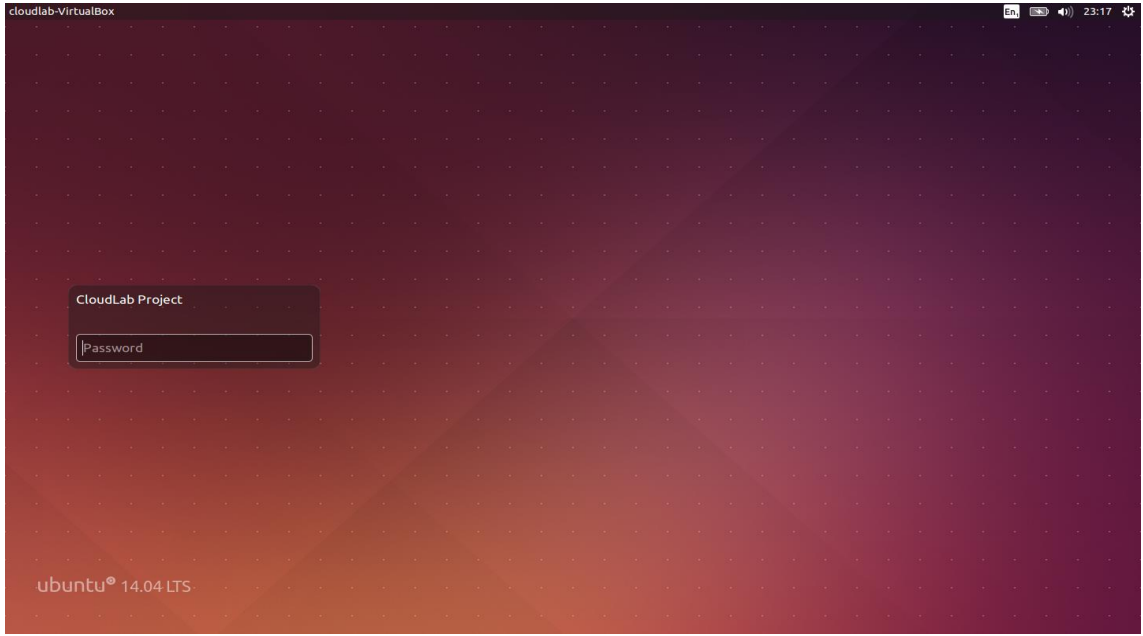

Figure 6: Ubuntu CloudLab Project Login Page

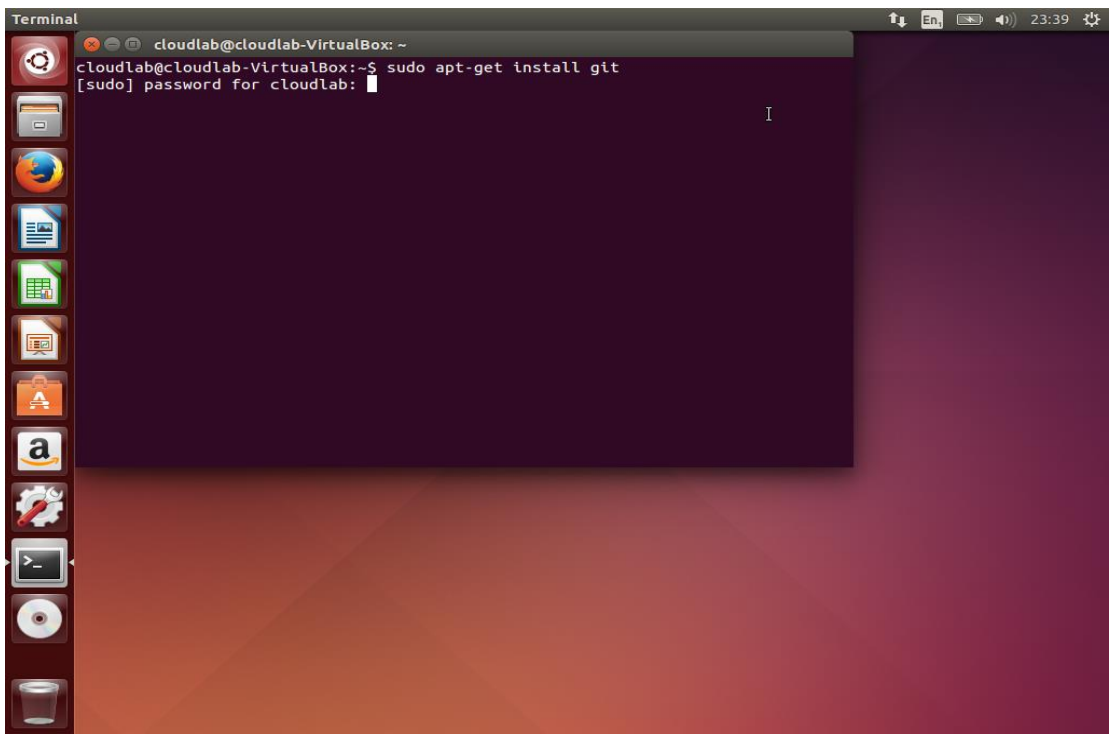

Figure 7: DevStack installation on Ubuntu

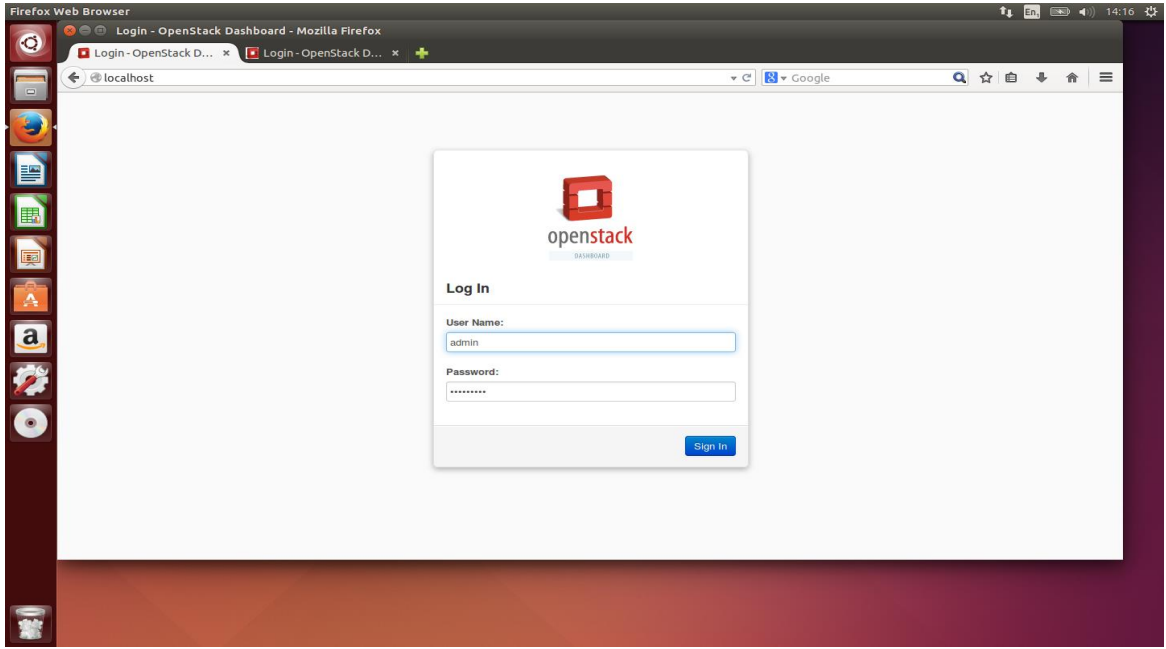

Figure 8: Open Stack Dashboard Login Page

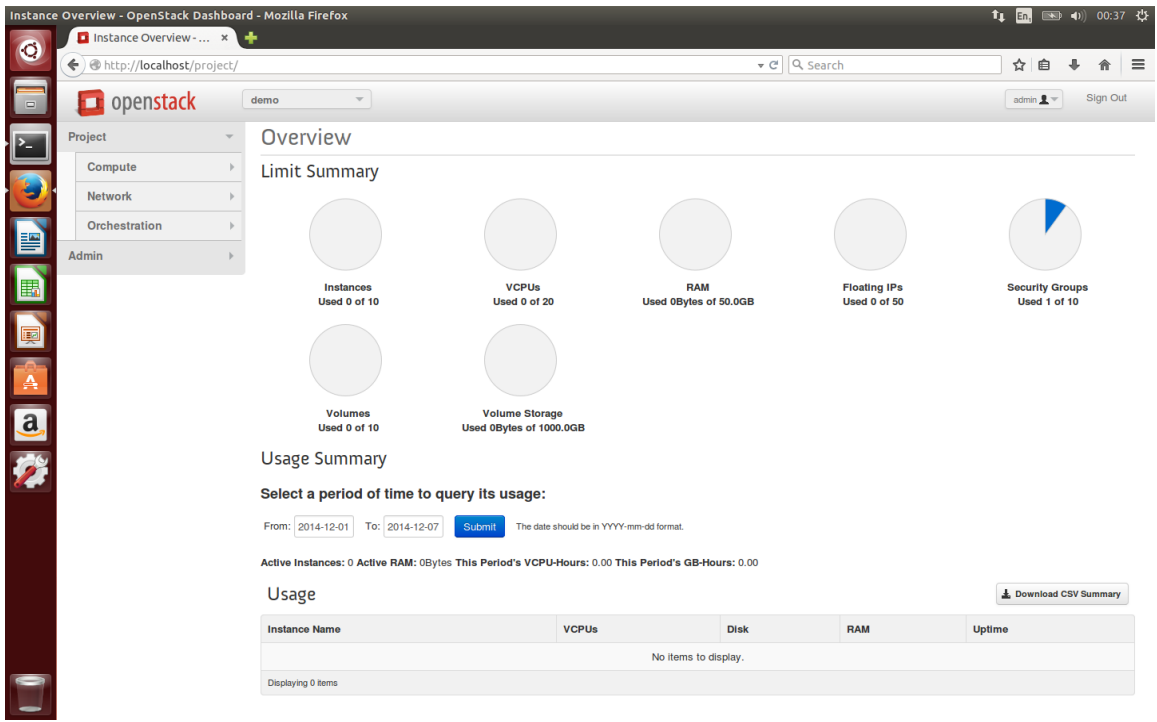

Figure 9: Open Stack Dashboard Overview

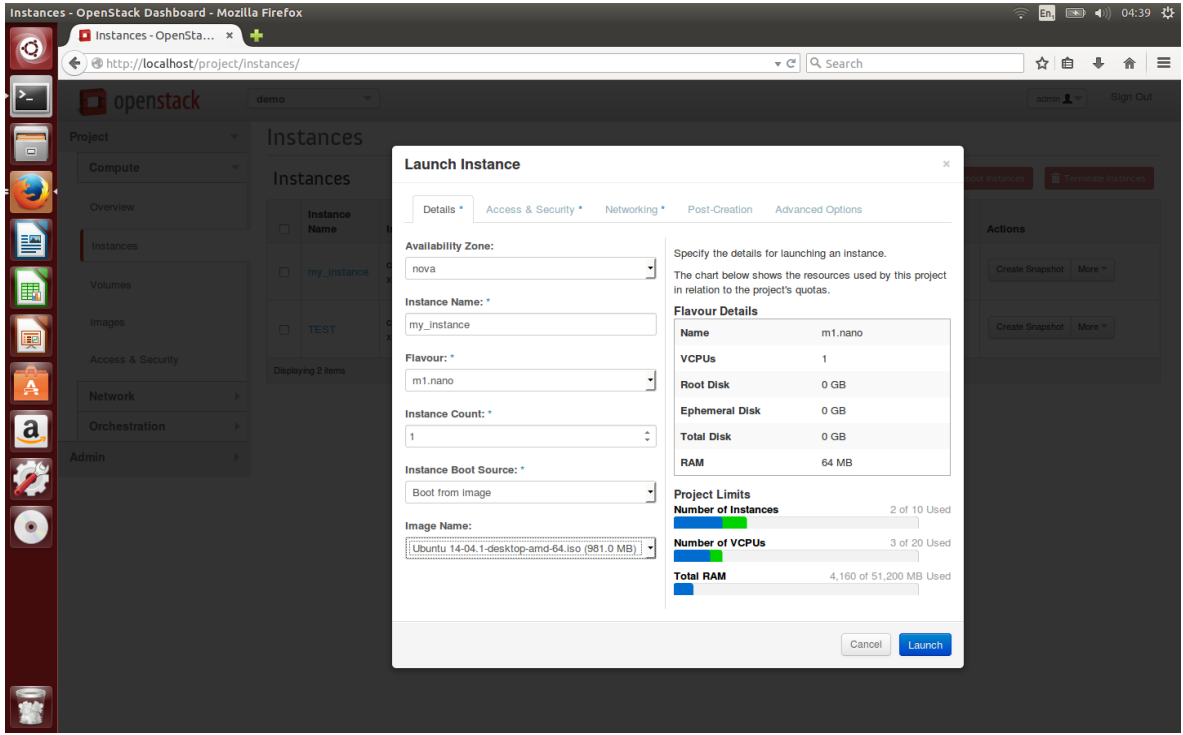

Figure 10: Creating an Instance on OpenStack

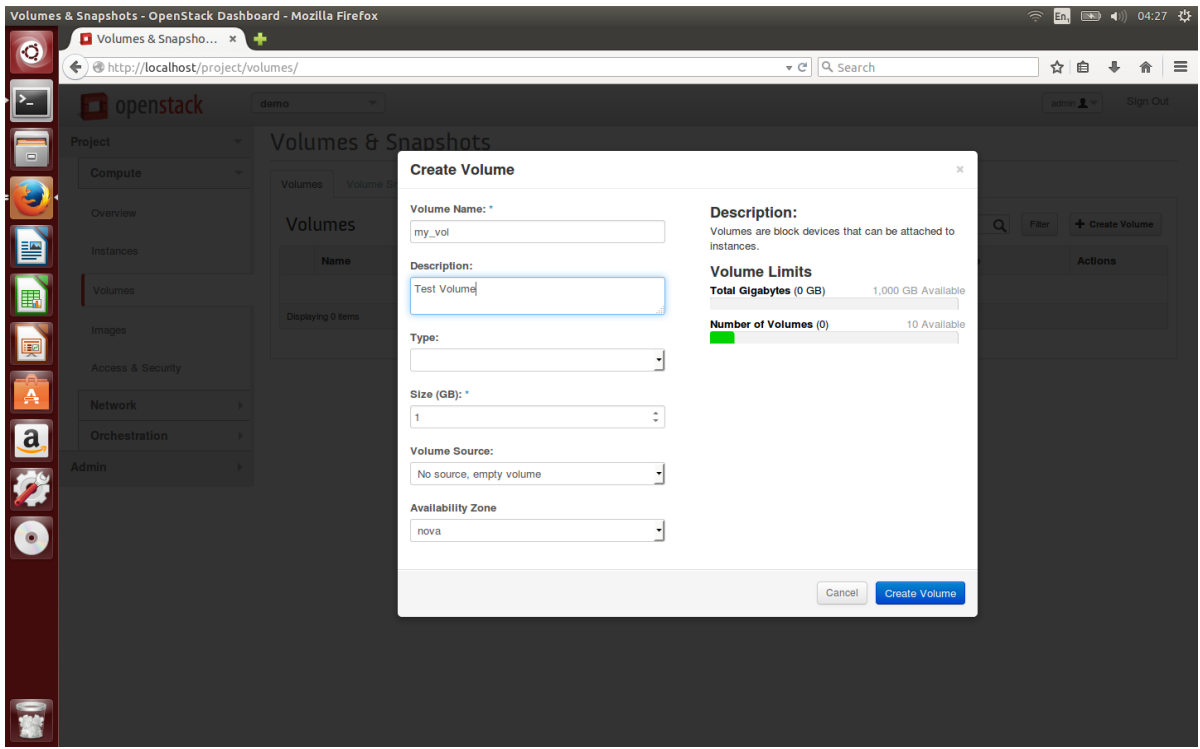

Figure 11: Creating a Volume on Openstack

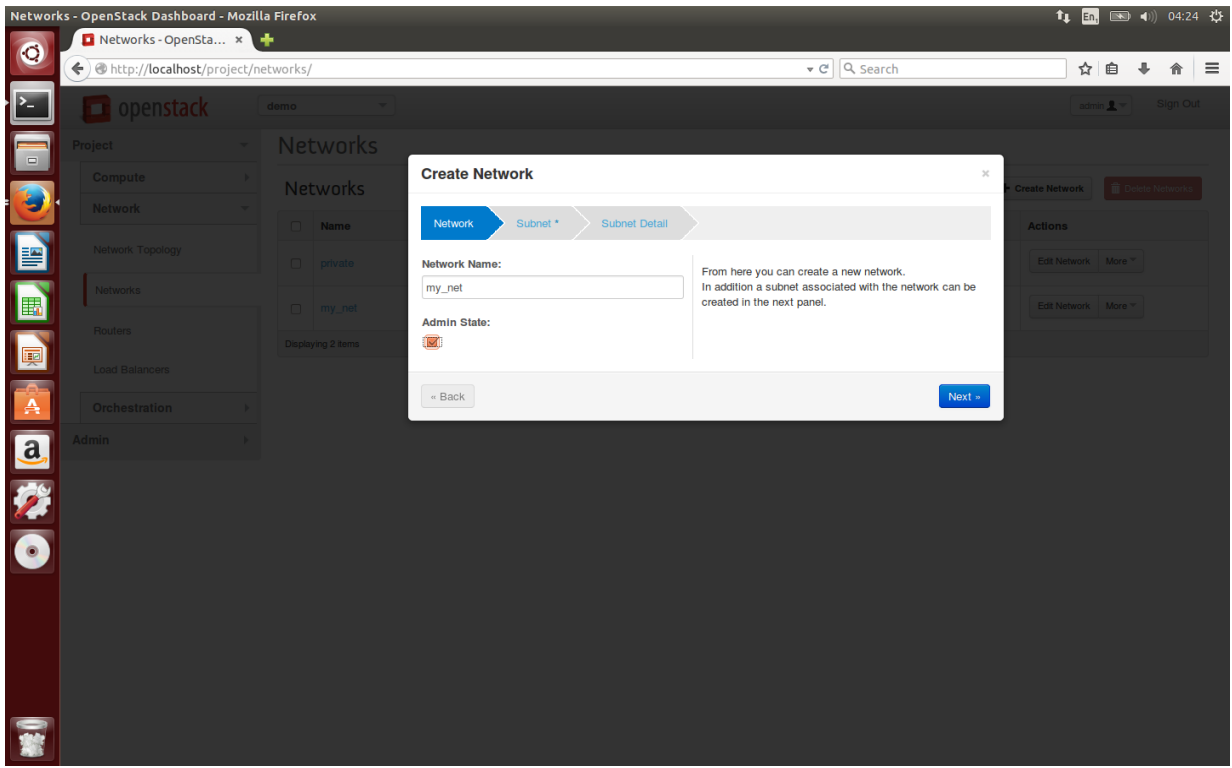

Figure 12: Creating a Network on Openstack

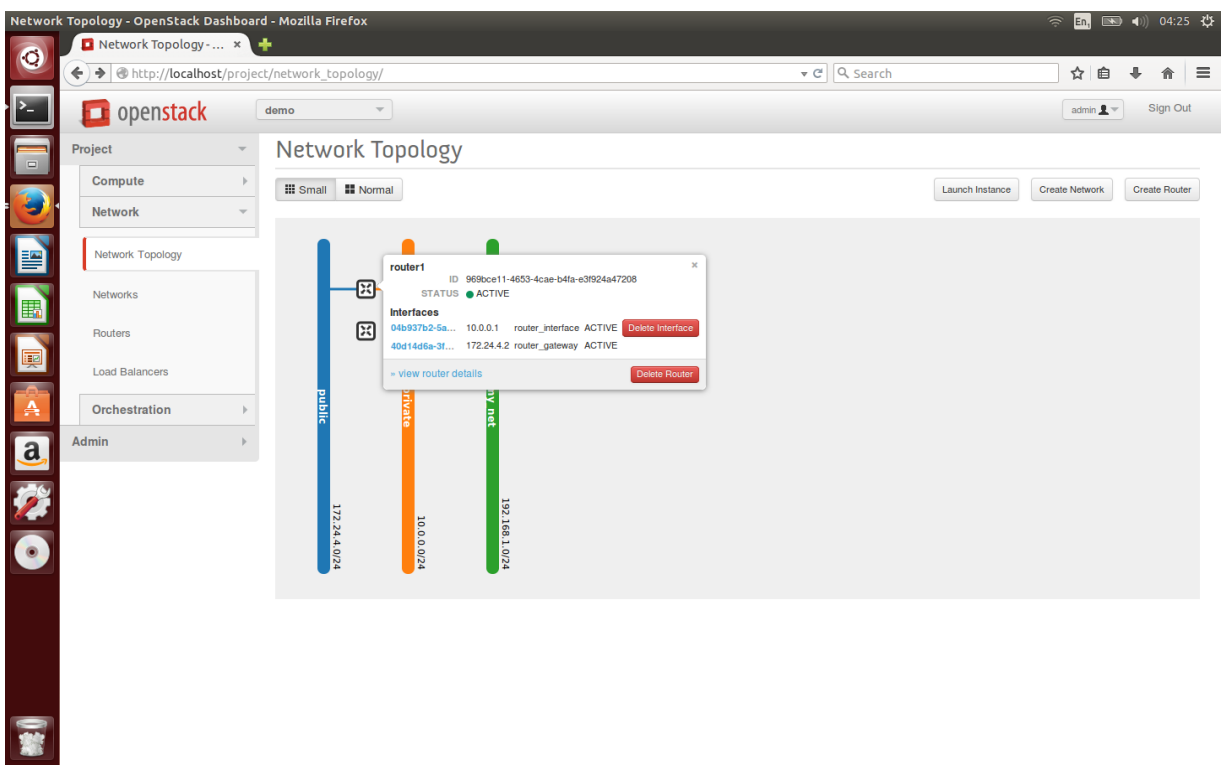

Figure13: Network Topology on Openstack

# Results

This experiment was carried out to show a proof of concept, highlighting steps involved in setting up a cloud lab for testing and experiments and to also demonstrate the efficiency, benefits and gain of adapting Cloud migration technology. Our approach was tested using three physical servers as shown in figure 3. In the setup, the first server hosts a web service and the second server hosts a database service as illustrated in figure 14. These exact physical servers, network equipment and their configurations were replaced by spinning up virtual instances running on the testbed Cloud.

Figure14 shows an architecture of our approach of the Cloud based migration solution with its network topology.

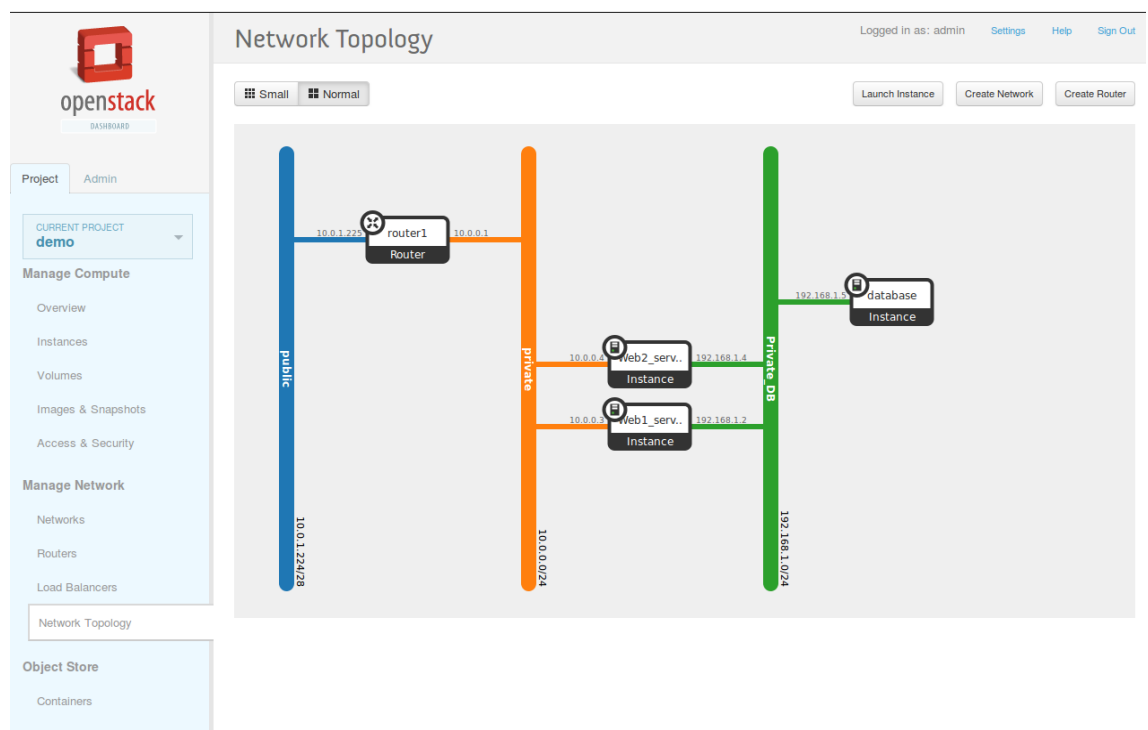

Figure 14: Cloud based migration solution Architecture

The gain of the approach is immediately noticeable. The implementation of the system requires only the creation, execution and spinning up of virtual machine instances by specifying the details and required specification in openstack as demonstrated in figure 10. Any direct manipulation of the physical servers or hardware equipment is not necessary. We also set up a load balancing solution by building a pool using the Round Robin method on HTTP protocol, all running without adding any additional physical hardware[3]. We then created a Virtual IP (VIP) address that can freely float between cluster nodes for the pool created.

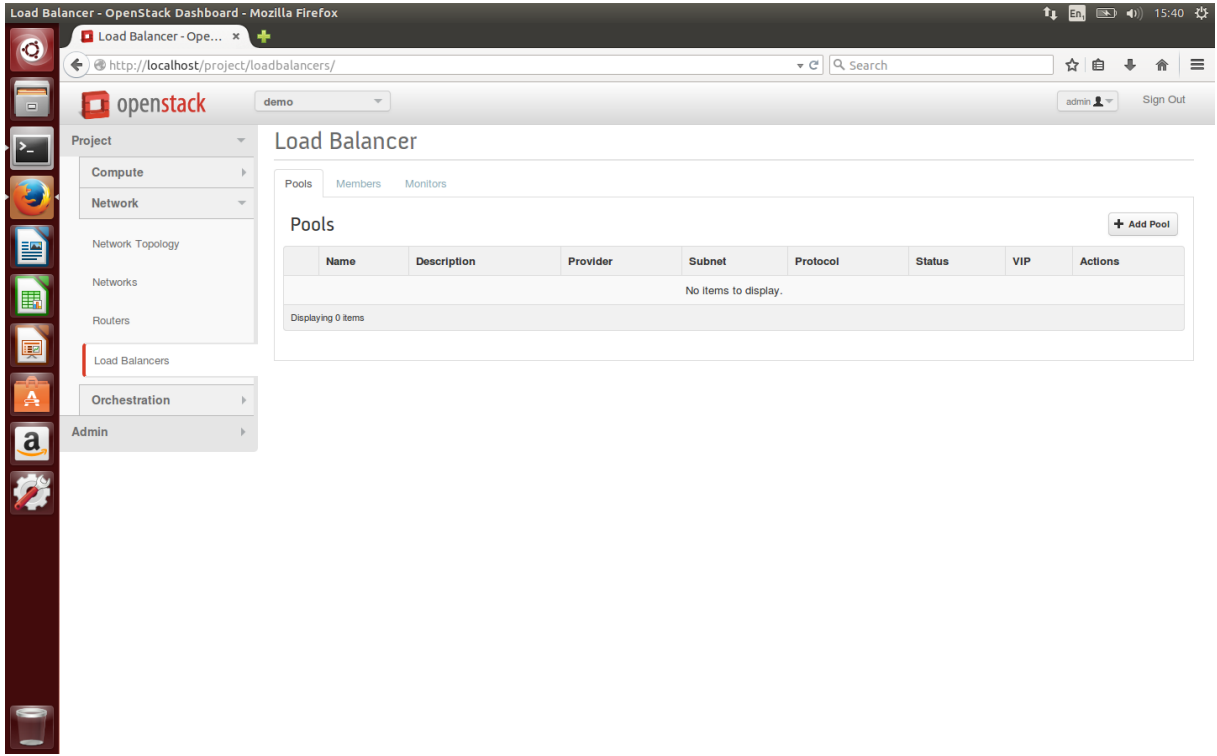

Figure 15: Load Balancing Dashboard

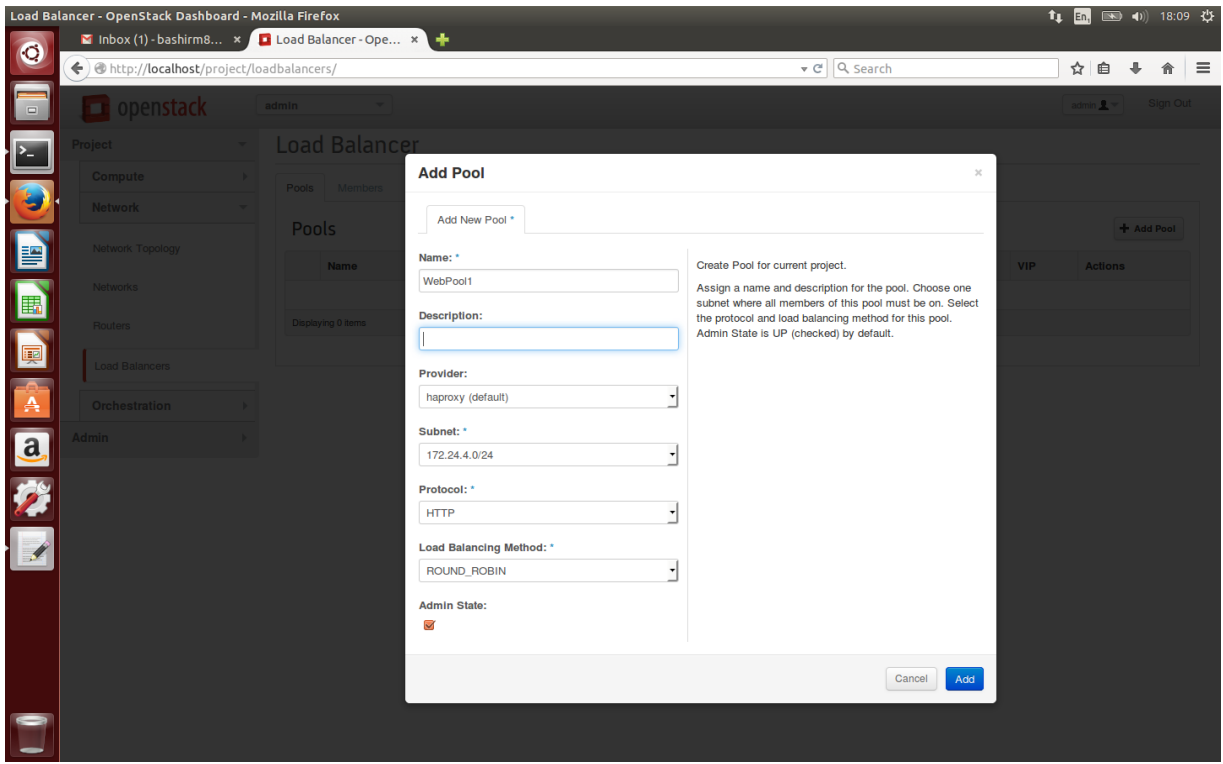

Figure 16: Creating a pool for Load balancing

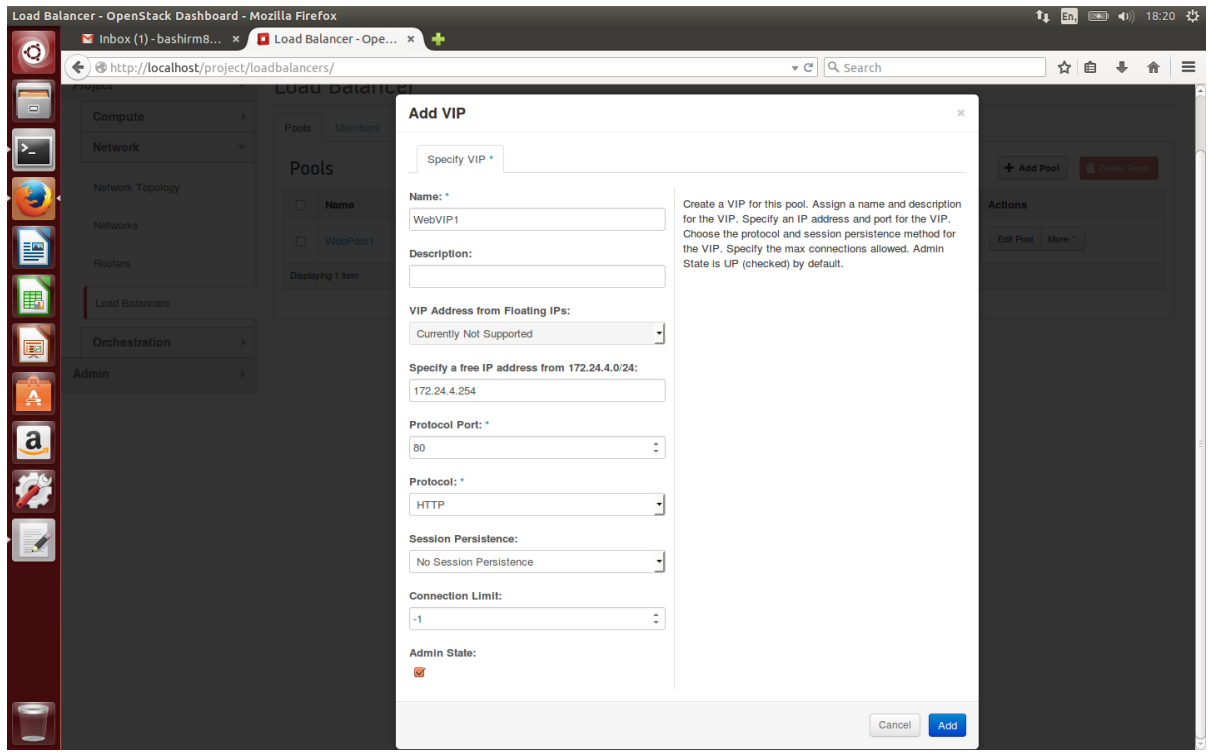

Figure 17: Creation of Virtual IP for the Pool

## Results & Observations

This section describes the results obtained:

1- The result of this study suggests that Devstack is a developer version and require execution of "$. /stack.sh" after every reboot which takes some time. The practical showed that there is no option available in devstack main trunk to avoid running stack.sh after every reboot. A faster and shorter way to avoid this is to always run "$. /rejoin-stack.sh".

2- During DevStack installation on a 32-bit CPU system: after running $./stack.sh, some errors were discovered which was as a result of processor compatibility issues. It was observed that Devstack can only run in its full capacity on a 64bit CPU system running on a 64/32-bit OS, but cannot run a 32-bit CPU system running on a 64/32-bit OS.

3- It was also observed that in order to have a successful installation without errors, the requirement is running a 64-bit OS so that we can have guests at the same level or at a lower level of 32-bit. A scenario where a 64-bit OS is trying to be launched on a 32-bit hypervisor is not acceptable. KVM supports 32-bit guests on 64-bit host, and any combination of Physical Address Extension (PAE) and not PAE guests and host.

## Conclusion & Recommendations

Growing popularity of cloud computing, especially using infrastructure as a service has inspired institutions to transform their existing infrastructure into a private or hybrid cloud. Even though openstack was not the first to propose open source cloud computing frameworks, but it works with popular enterprise and open source technologies making it ideal for heterogeneous infrastructure. OpenStack community of actively contributing developers is the largest among all open cloud solutions. Virtualization technology eliminates the capital costs by sharing and managing computing resources efficiently by scaling the infrastructure according to the service demands. But the use of virtualization setup may introduce substantial performance penalties in not carried out based on best practice. In this work, we studied the performance our experiment carried out to demonstrate the efficiency, benefits and gain of adapting Cloud migration technology. Physical servers were setup at the University of Bradford's Networking and computing lab to propose a case study to demonstrate the migrating feasibility from a classic web service solution to the Cloud. The gains and benefits obtained from this approach were highlighted.

As future work, we will continue using openstack through the analysis of Load-balance for client traffic from one network to application services, such as hosting and accessing virtual machines on the same and from a different network. Further research will also focus on fault tolerance techniques in cloud environments as mechanisms to prevent data loss and network outages costing the cloud computing environment.